\documentclass[12pt]{article}
\usepackage{amsmath,amssymb,amsthm,amsxtra,overpic,bbm,bm,epsfig,ulem}
\usepackage{color}
\usepackage{cite}

\usepackage{hyperref}
\usepackage{url}

\def\thefootnote{\fnsymbol{footnote}}

\begin{document}

\vspace{0.2cm}

\begin{center}
{\large\bf Identifying a minimal flavor symmetry of the seesaw mechanism \\
behind neutrino oscillations}
\end{center}

\vspace{0.2cm}

\begin{center}
{\bf Zhi-zhong Xing$^{1,2}$}
\footnote{E-mail: xingzz@ihep.ac.cn}
\\
{\small $^{1}$Institute of High Energy Physics and School of Physical Sciences, \\
University of Chinese Academy of Sciences, Beijing 100049, China \\
$^{2}$Center of High Energy Physics, Peking University, Beijing 100871, China}
\end{center}

\vspace{2cm}
\begin{abstract}
In the canonical seesaw framework flavor mixing and CP violation in weak
charged-current interactions of {\it light} and {\it heavy} Majorana
neutrinos are correlated with each other and described respectively by
the $3\times 3$ matrices $U$ and $R$. We show that the very possibility of
$\big|U^{}_{\mu i}\big| = \big|U^{}_{\tau i}\big|$ (for $i = 1, 2, 3$), which
is strongly indicated by current neutrino oscillation data, automatically leads to
a novel prediction $\big|R^{}_{\mu i}\big| = \big|R^{}_{\tau i}\big|$
(for $i = 1, 2, 3$). We prove that behind these two sets of equalities and the
experimental evidence for leptonic CP violation lies a minimal flavor
symmetry --- the overall neutrino mass term keeps invariant when
the left-handed neutrino fields transform as
$\nu^{}_{e \rm L} \to (\nu^{}_{e \rm L})^c$,
$\nu^{}_{\mu \rm L} \to (\nu^{}_{\tau \rm L})^c$,
$\nu^{}_{\tau \rm L} \to (\nu^{}_{\mu \rm L})^c$ and the right-handed
neutrino fields undergo an arbitrary unitary CP transformation.
Such a generalized $\mu$-$\tau$ reflection symmetry may help constrain
the flavor textures of active and sterile neutrinos to some extent in
the seesaw mechanism.
\end{abstract}

\newpage

\def\thefootnote{\arabic{footnote}}
\setcounter{footnote}{0}

\section{Motivation}

The discoveries of marvellous flavor oscillations in the atmospheric, solar, reactor
and accelerator neutrino (or antineutrino) experiments have constituted a
great success in particle physics~\cite{ParticleDataGroup:2020ssz}.
But the Standard Model (SM) of particle physics does not really offer an opportunity
for three neutrino flavors to oscillate from one type to another, simply because
it has required both neutrino masses and lepton flavor mixing to vanish.
It is therefore an absolute ``must" to introduce tiny neutrino masses and significant
lepton flavor mixing effects in a way that is beyond the SM framework, so as to fully
account for current experimental data on neutrino oscillations. In this connection,
however, the real theoretical challenges include how to pin down the true origin of
neutrino masses and single out the most likely flavor symmetry responsible for the
observed pattern of lepton flavor mixing and CP violation~\cite{Xing:2019vks}.

Regarding the issue of neutrino mass generation, it has commonly been accepted that
the tiny masses of three active neutrinos should be attributed to their Majorana
nature~\cite{Majorana:1937vz} and the existence of a few species of {\it sterile}
neutrinos which are much heavier than the Higgs boson~\cite{Minkowski:1977sc,Yanagida:1979as,
GellMann:1980vs,Glashow:1979nm,Mohapatra:1979ia,Schechter:1980gr,Schechter:1981cv}.
Such a popular {\it seesaw} mechanism works only in a qualitative
way, since the unknown flavor structures of active and sterile neutrinos unavoidably
forbid its quantitative predictability~\cite{Witten:2000dt}. Although quite a lot of
efforts have been made to constrain the neutrino textures with the help of various
flavor symmetry groups~\cite{Altarelli:2010gt,King:2013eh,Feruglio:2021sir},
which flavor symmetry really lies behind what we have observed remains unclear.
But a consensus seems to have essentially been reached: no matter how large
the true flavor symmetry group is, its {\it residual} symmetry at low energies
should serve as a {\it minimal} flavor symmetry in the neutrino sector.
Here the meaning of ``minimal" is two-fold. On the one hand, this flavor symmetry
can be described by one of the simplest discrete or continuous symmetry groups.
On the other hand, the pattern of lepton flavor mixing and CP violation determined
or constrained by such a simple flavor symmetry should be as close as possible to
the one extracted from the experimental data on neutrino oscillations.

Along this line of thought, we intend to identify a minimal flavor symmetry
associated with three species of active and sterile neutrinos in the canonical
seesaw mechanism by starting from the very possibility of
$\big|U^{}_{\mu i}\big| = \big|U^{}_{\tau i}\big|$
(for $i=1,2,3$) that is strongly favored by current neutrino oscillation data
for the $3\times 3$ Pontecorvo-Maki-Nakagawa-Sakata (PMNS) matrix
$U$~\cite{Pontecorvo:1957cp,Maki:1962mu,Pontecorvo:1967fh}.
Given the fact that $U$ is not exactly unitary but intrinsically
correlated with another $3\times 3$ flavor mixing matrix $R$ describing
the strength of weak charged-current interactions of heavy Majorana neutrinos via
the exact seesaw formula and the unitarity condition $UU^\dagger + RR^\dagger = I$,
we show that $\big|U^{}_{\mu i}\big| = \big|U^{}_{\tau i}\big|$ can automatically
lead to a novel prediction $\big|R^{}_{\mu i}\big| = \big|R^{}_{\tau i}\big|$
(for $i = 1, 2, 3$). We proceed to prove that behind these two sets of equalities
and the preliminary experimental evidence for leptonic CP violation~\cite{T2K:2019bcf}
lies an expected minimal flavor symmetry; namely, the overall neutrino mass term keeps
invariant when the left-handed neutrino fields transform as
$\nu^{}_{e \rm L} \to (\nu^{}_{e \rm L})^c$,
$\nu^{}_{\mu \rm L} \to (\nu^{}_{\tau \rm L})^c$,
$\nu^{}_{\tau \rm L} \to (\nu^{}_{\mu \rm L})^c$ and the right-handed
neutrino fields undergo an arbitrary unitary CP transformation.
With the help of a full Euler-like parametrization of the PMNS matrix $U$ and its 
counterpart $R$, we demonstrate that such a generalized $\mu$-$\tau$ reflection 
symmetry may help constrain the flavor structures of active and sterile neutrinos 
to some extent. So it should be able to enhance  
both predictability and testability of the canonical seesaw mechanism.

\setcounter{equation}{0}
\section{Proofs}

\subsection{The seesaw mechanism}

The trivial reason for vanishing neutrino masses in the SM is an absence of
the right-handed neutrino fields and thus an absence of the Yukawa interactions
between the Higgs and neutrino fields. It is therefore natural to minimally
extend the SM by adding three right-handed neutrino fields $N^{}_{\alpha \rm R}$
(for $\alpha = e, \mu, \tau$) and allowing for lepton number
violation~\cite{Minkowski:1977sc,Yanagida:1979as,GellMann:1980vs,Glashow:1979nm,
Mohapatra:1979ia,Schechter:1980gr,Schechter:1981cv}. In this case the gauge- and
Lorentz-invariant neutrino mass terms can be written as
\begin{eqnarray}
-{\cal L}^{}_\nu = \overline{\ell^{}_{\rm L}} \hspace{0.05cm} Y^{}_\nu
\widetilde{H} N^{}_{\rm R} + \frac{1}{2} \hspace{0.05cm} \overline{(N^{}_{\rm R})^c}
\hspace{0.05cm} M^{}_{\rm R} N^{}_{\rm R} + {\rm h.c.} \; ,
\label{2.1}
%     (2.1)
\end{eqnarray}
where $\ell^{}_{\rm L}$ denotes the $\rm SU(2)^{}_{\rm L}$ doublet of the left-handed
lepton fields, $\widetilde{H} \equiv {\rm i} \sigma^{}_2 H^*$ with
$H$ being the Higgs doublet and $\sigma^{}_2$ being the second Pauli matrix,
$N^{}_{\rm R} = (N^{}_{e \rm R} , N^{}_{\mu \rm R} , N^{}_{\tau \rm R})^T$ is the column
vector of three right-handed neutrino fields which are the $\rm SU(2)^{}_{\rm L}$
singlets, $(N^{}_{\rm R})^c \equiv {\cal C} \overline{N^{}_{\rm R}}^T$ with $\cal C$
being the charge-conjugation operator and satisfying
${\cal C} \gamma^T_\mu {\cal C}^{-1} = -\gamma^{}_\mu$ and ${\cal C}^{-1} =
{\cal C}^\dagger = {\cal C}^T = -{\cal C}$, $Y^{}_\nu$ represents an arbitrary $3\times 3$
Yukawa coupling matrix, and $M^{}_{\rm R}$ stands for a symmetric $3\times 3$
Majorana mass matrix. After spontaneous electroweak gauge symmetry breaking,
Eq.~(\ref{2.1}) becomes
\begin{eqnarray}
-{\cal L}^\prime_\nu = \frac{1}{2} \hspace{0.05cm} \overline{\left[
\nu^{}_{\rm L} \hspace{0.2cm} (N^{}_{\rm R})^c\right]}
\left(\begin{matrix} {\bf 0} & M^{}_{\rm D} \cr
M^T_{\rm D} & M^{}_{\rm R} \end{matrix}\right)
\left[\begin{matrix} (\nu^{}_{\rm L})^c \cr N^{}_{\rm R} \end{matrix}
\right] + {\rm h.c.} \; ,
\label{2.2}
%     (2.2)
\end{eqnarray}
where $\nu^{}_{\rm L} = (\nu^{}_{e \rm L}, \nu^{}_{\mu \rm L}, \nu^{}_{\tau \rm L})^T$
denotes the column vector of three left-handed neutrino fields,
$M^{}_{\rm D} \equiv Y^{}_\nu \langle H\rangle$ with $\langle H\rangle$ being the
vacuum expectation value of the Higgs field, and $\bf 0$ denotes
the $3\times 3$ zero matrix. Note that $M^{}_{\rm D}$
is in general neither Hermitian nor symmetric. The overall symmetric $6\times 6$
neutrino mass matrix in Eq.~(\ref{2.2}) can be diagonalized by the
unitary transformation
\begin{eqnarray}
\left( \begin{matrix} U & R \cr S & Q \cr \end{matrix}
\right)^{\hspace{-0.05cm} \dagger} \left ( \begin{matrix} {\bf 0} & M^{}_{\rm D}
\cr M^{T}_{\rm D} & M^{}_{\rm R} \cr \end{matrix} \right ) \left(
\begin{matrix} U & R \cr S & Q \cr \end{matrix} \right)^{\hspace{-0.05cm} *}
= \left( \begin{matrix} D^{}_\nu & {\bf 0} \cr {\bf 0} &
D^{}_N \cr \end{matrix} \right) \; ,
\label{2.3}
%     (2.3)
\end{eqnarray}
where $D^{}_\nu \equiv {\rm Diag}\big\{m^{}_1, m^{}_2, m^{}_3\big\}$ and
$D^{}_N \equiv {\rm Diag}\big\{M^{}_1, M^{}_2, M^{}_3 \big\}$ with $m^{}_i$ and
$M^{}_i$ (for $i = 1, 2, 3$) being the masses of active and sterile Majorana neutrinos,
respectively. The $3\times 3$ submatrices $U$, $R$, $S$ and $Q$ in Eq.~(\ref{2.3})
satisfy the unitarity conditions:
\begin{eqnarray}
U U^\dagger + RR^\dagger = SS^\dagger + Q Q^\dagger
\hspace{-0.2cm} & = & \hspace{-0.2cm} I \; ,
\nonumber \\
U^\dagger U + S^\dagger S = R^\dagger R + Q^\dagger Q
\hspace{-0.2cm} & = & \hspace{-0.2cm} I \; ,
\nonumber \\
U S^\dagger + R Q^\dagger = U^\dagger R + S^\dagger Q
\hspace{-0.2cm} & = & \hspace{-0.2cm} {\bf 0} \; ; 
\label{2.4}
%     (2.4)
\end{eqnarray}
and the exact {\it seesaw} formula that characterizes a kind of balance between the
light and heavy neutrino sectors can also be obtained from 
Eq.~(\ref{2.3})~\cite{Xing:2009ce}:
\begin{eqnarray}
U D^{}_\nu U^T + R D^{}_N R^T = {\bf 0} \; .
\label{2.5}
%     (2.5)
\end{eqnarray}
So the smallness of $m^{}_i$ is essentially ascribed to the highly suppressed
magnitude of $R$ which signifies the largeness of $M^{}_i$ with respect to the electroweak
scale (i.e., $R \sim {\cal O}(M^{}_{\rm D}/M^{}_{\rm R}) \ll 1$ in the leading-order
approximation~\cite{Minkowski:1977sc,Yanagida:1979as,GellMann:1980vs,Glashow:1979nm,
Mohapatra:1979ia,Schechter:1980gr,Schechter:1981cv}).
In this canonical seesaw framework $\nu^{}_{\alpha}$
(for $\alpha = e, \mu, \tau$) can be expressed as a linear combination of
the mass eigenstates of three active (light) neutrinos and three sterile (heavy)
neutrinos (i.e., $\nu^{}_{i} = (\nu^{}_i)^c$ and
$N^{}_{i} = (N^{}_i)^c$ for $i = 1, 2, 3$). The standard weak charged-current
interactions of these six Majorana neutrinos is therefore given by
\begin{eqnarray}
-{\cal L}^{}_{\rm cc} = \frac{g}{\sqrt{2}} \hspace{0.1cm}
\overline{\left(e \hspace{0.2cm} \mu \hspace{0.2cm}
\tau \right)^{}_{\rm L}} \hspace{0.1cm} \gamma^\mu
\left[ U \left( \begin{matrix} \nu^{}_{1} \cr \nu^{}_{2} \cr
\nu^{}_{3} \cr\end{matrix} \right)^{}_{\hspace{-0.08cm} \rm L} + R \left(
\begin{matrix} N^{}_{1} \cr N^{}_{2} \cr N^{}_{3}
\cr\end{matrix}\right)^{}_{\hspace{-0.08cm} \rm L} \right] W^-_\mu + {\rm h.c.} \; .
\label{2.6}
%     (2.6)
\end{eqnarray}
It is obvious that the PMNS matrix $U$ describes flavor mixing and CP violation
of three active neutrinos (e.g., in neutrino oscillations), and its
counterpart $R$ measures the strength of weak charged-current interactions of
three sterile neutrinos (e.g., in precision collider physics).
The validity of such a seesaw mechanism means that $R \neq {\bf 0}$ must hold
no matter how small its nine elements may be, otherwise the active neutrinos
would have no chance to acquire their masses from Eqs.~(\ref{2.1}) and (\ref{2.5}).
As a consequence of $UU^\dagger = I - RR^\dagger \neq I$, the PMNS
matrix $U$ must be non-unitary in the seesaw framework although its deviation
from unitarity is expected to be very small. A careful analysis of current
electroweak precision measurements and neutrino oscillation data has constrained
$U$ to be unitary at the ${\cal O}(10^{-2})$ sensitivity
level~\cite{Antusch:2006vwa,Antusch:2009gn,Blennow:2016jkn,Hu:2020oba,Wang:2021rsi}.
In other words, possible non-unitarity effects hiding in $U$ are expected to be at
most of ${\cal O}(10^{-2})$ and hence cannot be identified up to the accuracy
level of today's neutrino oscillation experiments.

\subsection{Implications of $\big|U^{}_{\mu i}\big| = \big|U^{}_{\tau i}\big|$}

\begin{figure}[t]
\begin{center}
\includegraphics[width=2.9in]{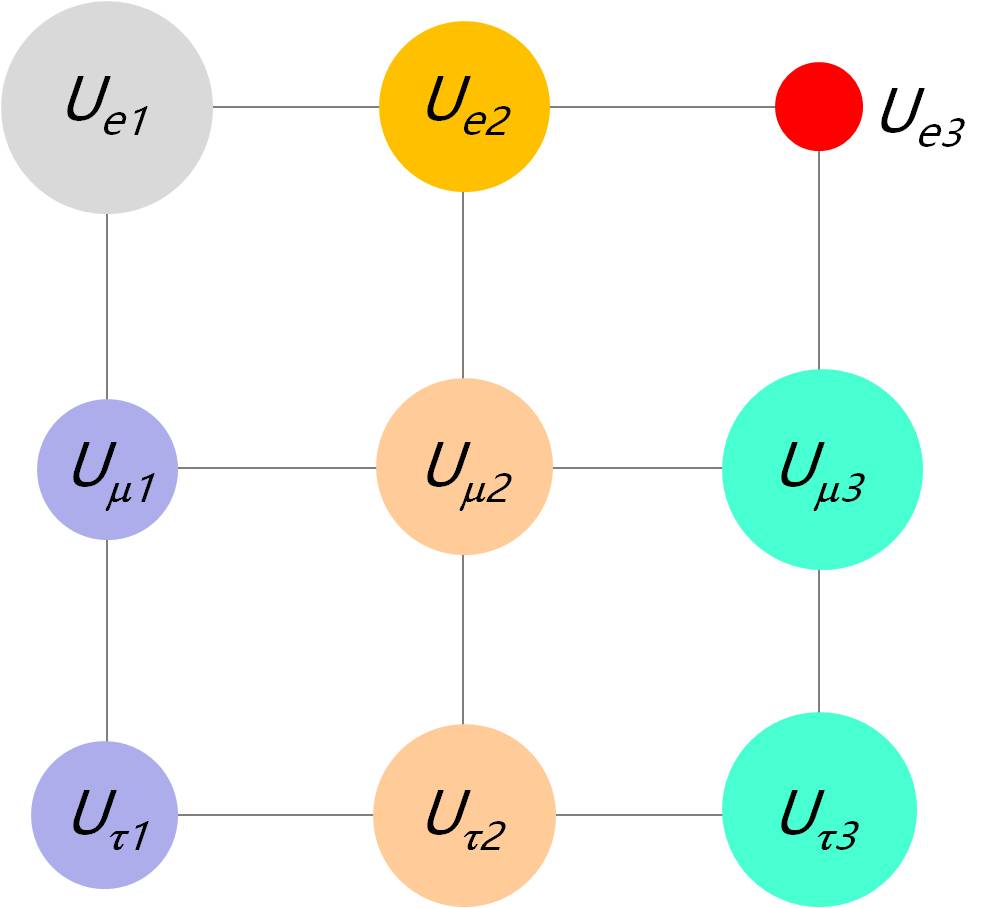}
\vspace{-0.1cm}
\caption{An illustration of the relative moduli of nine PMNS matrix
elements, where the colored area of each circle is proportional to
$\big|U^{}_{\alpha i}\big|$ (for $\alpha = e, \mu, \tau$ and $i = 1, 2, 3$) on
the same scaling.}
\label{Fig:1}
\end{center}
\end{figure}

A global analysis of the latest atmospheric, solar, reactor and accelerator
neutrino oscillation data yields the $3\sigma$ intervals of nine elements of
the PMNS matrix $U$ as follows~\cite{Esteban:2020cvm,Capozzi:2021fjo}:
\begin{eqnarray}
&& \big|U^{}_{e 1}\big| = 0.801 \to 0.845 \; , \hspace{0.03cm}
\quad
\big|U^{}_{e 2}\big| = 0.513 \to 0.579 \; , \hspace{0.04cm}
\quad
\big|U^{}_{e 3}\big| = 0.143 \to 0.155 \; , \hspace{0.9cm}
\nonumber \\
&& \big|U^{}_{\mu 1}\big| = 0.234 \to 0.500 \; ,
\quad
\big|U^{}_{\mu 2}\big| = 0.471 \to 0.689 \; ,
\quad
\big|U^{}_{\mu 3}\big| = 0.637 \to 0.776 \; ,
\nonumber \\
&& \big|U^{}_{\tau 1}\big| = 0.271 \to 0.525 \; , \hspace{0.01cm}
\quad
\big|U^{}_{\tau 2}\big| = 0.477 \to 0.694 \; , \hspace{0.01cm}
\quad
\big|U^{}_{\tau 3}\big| = 0.613 \to 0.756 \; .
\label{2.7}
%     (2.7)
\end{eqnarray}
As argued above, these numerical results are actually insensitive to small
unitarity violation of $U$ although they were originally obtained by assuming
$U$ to be unitary. The relative magnitudes of $\big|U^{}_{\alpha i}\big|$ (for
$\alpha = e, \mu, \tau$ and $i = 1, 2, 3$) are more intuitively illustrated
in Fig.~\ref{Fig:1}, where the colored area of each circle is proportional to the
size of $\big|U^{}_{\alpha i}\big|$ on the same scaling. We see that the {\it very}
possibility of $\big|U^{}_{\mu i}\big| = \big|U^{}_{\tau i}\big|$, which is
strongly favored by current experimental data, constitutes the most salient feature
of $U$ known to us
%%%%%%%%%%%%%%%%%%%%%%%%%%%%%%%%%%%%%%%%%%%%%%%%%%%%%%%%%%%%%%%%%%%%%%%%%%%%%%%%%
\footnote{Note that a somewhat weaker possibility,
$\big|U^{}_{e 2}\big| = \big|U^{}_{\mu 2}\big| = \big|U^{}_{\tau 2}\big| = 1/\sqrt{3}$,
has also attracted some interest of model building~\cite{Harrison:2002et} and
is apparently consistent with the case of
$\big|U^{}_{\mu i}\big| = \big|U^{}_{\tau i}\big|$
(for $i = 1, 2, 3$) under discussion.}.
%%%%%%%%%%%%%%%%%%%%%%%%%%%%%%%%%%%%%%%%%%%%%%%%%%%%%%%%%%%%%%%%%%%%%%%%%%%%%%%%%
This observation motivates us to consider how $U^{}_{\mu i}$ is directly related
to $U^{}_{\tau i}$ or $U^*_{\tau i}$ in a given flavor basis and whether their
rephasing-dependent relationship is naturally attributed to an underlying flavor
symmetry. In the same flavor basis $U^{}_{e i}$ is expected to either stay unchanged
or become its complex conjugate, such that $\big|U^{}_{e i}\big|$ keeps invariant.
We find that there are two typical possibilities of this kind that allow the PMNS
matrix elements $U^{}_{e i}$, $U^{}_{\mu i}$ and $U^{}_{\tau i}$ to transform together.
%%%%%%%%%%%%%%%%%%%%%%%%%%%% Figure 1 %%%%%%%%%%%%%%%%%%%%%%%%%%%%%%%%%%%%%
%%%%%%%%%%%%%%%%%%%%%%%%%%%%%%%%%%%%%%%%%%%%%%%%%%%%%%%%%%%%%%%%%%%%%%%%%%%%

{\bf Case A:} $U = {\cal P} U^*$, where the real orthogonal ``$(\mu, \tau)$-permutation"
matrix $\cal P$ takes the form
\begin{eqnarray}
{\cal P} = {\cal P}^T = {\cal P}^\dagger =
\left(\begin{matrix} 1 & ~0~ & 0 \cr
0 & 0 & 1 \cr 0 & 1 & 0 \cr\end{matrix}\right) \; ,
\label{2.8}
%     (2.8)
\end{eqnarray}
is certainly a simple but nontrivial solution to
$\big|U^{}_{\mu i}\big| = \big|U^{}_{\tau i}\big|$. Substituting
$U = {\cal P} U^*$ into the exact seesaw formula in Eq.~(\ref{2.5})
and taking the complex conjugate for the whole equation, we obtain
\begin{eqnarray}
U D^{}_\nu U^T + {\cal P} R^* D^{}_N ({\cal P} R^*)^T = {\bf 0} \; .
\label{2.9}
%     (2.9)
\end{eqnarray}
Then a comparison between Eqs.~(\ref{2.5}) and (\ref{2.9}) leads us to 
$R = {\cal P} R^*$,
This observation implies that the novel and rephasing-invariant prediction
$\big|R^{}_{\mu i}\big| = \big|R^{}_{\tau i}\big|$ is actually a natural
consequence of $\big|U^{}_{\mu i}\big| = \big|U^{}_{\tau i}\big|$ in
the canonical seesaw mechanism. Now let us proceed to substitute
$U = {\cal P} U^*$ and $R = {\cal P} R^*$ into the unitarity conditions
listed in Eq.~(\ref{2.4}). After a careful but straightforward check, we find that
$S = {\cal T} S^*$ and $Q = {\cal T} Q^*$ with $\cal T$ being an arbitrary
unitary matrix satisfy all the normalization and orthogonality relations
in Eq.~(\ref{2.4}). In short, we have
\begin{eqnarray}
\underbrace{U = {\cal P} U^* ~\longrightarrow~ R = {\cal P} R^*} \hspace{0.2cm}
\nonumber \\
\Downarrow \hspace{2.3cm}
\nonumber \\
\overbrace{S = {\cal T} S^* \hspace{1.26cm} Q = {\cal T} Q^*} \hspace{0.24cm}
\label{2.10}
%     (2.10)
\end{eqnarray}
as constrained by Eqs.~(\ref{2.4}) and (\ref{2.5}). One will see that these relations
can help fix the flavor textures of active and sterile Majorana neutrinos
to a large extent.

\begin{figure}[t]
\begin{center}
\includegraphics[width=3.8in]{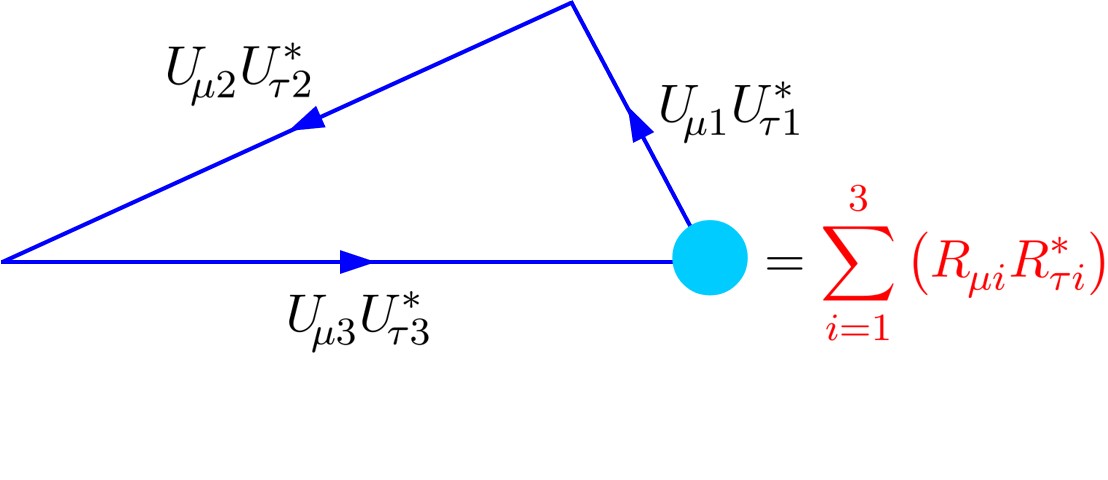}
\vspace{-1.3cm}
\caption{An illustration of the unitarity polygon defined by the
$(\mu,\tau)$-associated orthogonality condition in the complex plane,
where the {\it effective} vertex in sky blue denotes small corrections of
the active-sterile flavor mixing effects to the three-flavor unitarity triangle
in the seesaw framework.}
\label{Fig:2}
\end{center}
\end{figure}

{\bf Case B:} $U = {\cal P} U \zeta$, where $\zeta = {\rm Diag}\big\{
\eta^{}_1 , \eta^{}_2 , \eta^{}_3\big\}$ with $\eta^{}_i = \pm 1$,
is the other simple but typical solution to
$\big|U^{}_{\mu i}\big| = \big|U^{}_{\tau i}\big|$ (for $i = 1, 2, 3$).
Note that replacing $\zeta$ with a diagonal phase matrix is also
allowed in this regard, but such a possibility will be disfavored when
the corresponding expression of $U$ is substituted into the exact seesaw
relation in Eq.~(\ref{2.5}). In this case one may start from
$U = {\cal P} U \zeta$ to similarly prove that
$R = {\cal P} R \zeta^\prime$, $S = {\cal T}^\prime S \zeta$ and
$Q = {\cal T}^\prime Q \zeta^\prime$ hold with the help of Eqs.~(\ref{2.4}) 
and (\ref{2.5}),
where $\zeta^\prime = {\rm Diag}\big\{\eta^{\prime}_1 , \eta^{\prime}_2 ,
\eta^{\prime}_3\big\}$ (for $\eta^{\prime}_i = \pm 1$) and ${\cal T}^\prime$
is another arbitrary unitary matrix. Concentrating on the
$(\mu, \tau)$-associated orthogonality condition, we obtain
\begin{eqnarray}
\sum^3_{i=1} \left(U^{}_{\mu i} U^*_{\tau i} + R^{}_{\mu i} R^*_{\tau i}\right)
= \sum^3_{i=1} \left(\eta^{}_i \big|U^{}_{\mu i}\big|^2 +
\eta^\prime_i \big|R^{}_{\mu i}\big|^2 \right) = 0 \; ,
\label{2.11}
%     (2.11)
\end{eqnarray}
where the magnitude of $\big|R^{}_{\mu i}\big|^2$ is expected to be of
or below ${\cal O}(10^{-2})$. Fig.~\ref{Fig:2} is an illustration
of the unitarity polygon defined by the above orthogonality relation in
the complex plane. It becomes clear that the possibility of
$\eta^{}_1 = \eta^{}_2 = \eta^{}_3$ has to be abandoned, otherwise Eq.~(\ref{2.11})
would be in conflict with the corresponding normalization condition of
$U^{}_{\mu i}$ and $R^{}_{\mu i}$ (for $i = 1, 2, 3$).
Given the fact that $\big| U^{}_{\mu 3}\big|$ is definitely greater than
$\big| U^{}_{\mu 1}\big|$ and most likely greater than $\big| U^{}_{\mu 2}\big|$,
as one can see from Eq.~(\ref{2.7}), we find that the choice of
$\eta^{}_1 = \eta^{}_2 = -\eta^{}_3$ is interesting because it results in
\begin{eqnarray}
\big|U^{}_{\mu 1}\big|^2 + \big|U^{}_{\mu 2}\big|^2 \hspace{-0.2cm} & = &
\hspace{-0.2cm} \frac{1}{2} \left[ 1 - \sum^3_{i=1} \left(1 + \eta^{}_1
\eta^\prime_i \right) \big|R^{}_{\mu i}\big|^2\right] \; ,
\nonumber \\
\big|U^{}_{\tau 3}\big|^2 = \big|U^{}_{\mu 3}\big|^2 \hspace{-0.2cm} & = &
\hspace{-0.2cm} \frac{1}{2} \left[ 1 - \sum^3_{i=1} \left(1 - \eta^{}_1
\eta^\prime_i \right) \big|R^{}_{\mu i}\big|^2\right] \; . 
\label{2.12}
%     (2.12)
\end{eqnarray}
However, this possibility indicates that $\big|U^{}_{e 3}\big|$ must be remarkably
smaller than its true value (i.e., $\big|U^{}_{e 3}\big| \simeq 0.15$) that
has been measured by the Daya Bay Collaboration~\cite{DayaBay:2012fng}.
On the other hand, a highly suppressed value of $\big|U^{}_{e 3}\big|$ would
give rise to a strong suppression of leptonic CP violation in neutrino
oscillations and thus contradict the $3\sigma$ evidence for CP violation
established recently by the T2K Collaboration~\cite{T2K:2019bcf}.
So our subsequent discussions will focus only on Case A, which is more likely
to lead us to a {\it minimal} flavor symmetry of the seesaw mechanism.
%%%%%%%%%%%%%%%%%%%%%%%%%%%% Figure 2 %%%%%%%%%%%%%%%%%%%%%%%%%%%%%%%%%%%%%
%%%%%%%%%%%%%%%%%%%%%%%%%%%%%%%%%%%%%%%%%%%%%%%%%%%%%%%%%%%%%%%%%%%%%%%%%%%%

\subsection{A minimal flavor symmetry}

Now that the rephasing-dependent relation $U = {\cal P} U^*$ {\it does} assure
that the rephasing-independent equalities
$\big|U^{}_{\mu i}\big| = \big|U^{}_{\tau i}\big|$ and
$\big|R^{}_{\mu i}\big| = \big|R^{}_{\tau i}\big|$ hold, it is very likely to
hint at a kind of flavor symmetry of the seesaw mechanism which just lies behind
the observed pattern of $U$. To pin down such a possible flavor symmetry, let
us first substitute $U = {\cal P} U^*$ and its counterparts
$R = {\cal P} R^*$, $S = {\cal T} S^*$ and $Q = {\cal T} Q^*$ into Eq.~(\ref{2.3}) 
and then take the complex conjugate for the whole equation. In this way we are
immediately left with
\begin{eqnarray}
\left( \begin{matrix} U & R \cr S & Q \cr \end{matrix}
\right)^{\hspace{-0.05cm} \dagger} \left ( \begin{matrix} {\bf 0}
& {\cal P} M^{*}_{\rm D} {\cal T}
\cr {\cal T}^T M^{\dagger}_{\rm D} {\cal P}
& {\cal T}^T M^{*}_{\rm R} {\cal T} \cr \end{matrix} \right ) \left(
\begin{matrix} U & R \cr S & Q \cr \end{matrix} \right)^{\hspace{-0.05cm} *}
= \left( \begin{matrix} D^{}_\nu & {\bf 0} \cr {\bf 0} &
D^{}_N \cr \end{matrix} \right) \; .
\label{2.13}
%     (2.13)
\end{eqnarray}
A comparison between Eqs.~(\ref{2.3}) and (\ref{2.13}) leads us to the following
constraint equations:
\begin{eqnarray}
M^{}_{\rm D} = {\cal P} M^{*}_{\rm D} {\cal T} \; , \quad
M^{}_{\rm R} = {\cal T}^T M^{*}_{\rm R} {\cal T} \; .
\label{2.14}
%     (2.14)
\end{eqnarray}
This new result is in principle powerful to constrain the flavor textures
of $M^{}_{\rm D}$ and $M^{}_{\rm R}$, but its application is in practice
limited by the fact that $\cal T$ is an arbitrary unitary matrix. For this
reason the assumption of a specific form of ${\cal T}$ actually corresponds
to a specific seesaw model with the $\mu$-$\tau$ reflection symmetry. Here
let us consider two simple but instructive scenarios of $\cal T$ for example.

{\bf Scenario A:} ${\cal T} = I$. In this simplest case $M^{}_{\rm R} = M^*_{\rm R}$
is a real symmetric matrix and can always be arranged to be diagonal in the
beginning (i.e., $M^{}_{\rm R} = D^{}_N$) by taking a proper basis for the
right-handed neutrino fields in Eq. (\ref{2.1}), and 
$M^{}_{\rm D} = {\cal P} M^*_{\rm D}$ is constrained as
\begin{eqnarray}
M^{}_{\rm D} \hspace{-0.2cm} & = & \hspace{-0.2cm}
\left(\begin{matrix} X^{}_{11} & X^{}_{12} & X^{}_{13} \cr
X^{}_{21} & X^{}_{22} & X^{}_{23} \cr
X^*_{21} & X^*_{22} & X^*_{23} \end{matrix}\right) \; ,
\label{2.15}
%     (2.15)
\end{eqnarray}
where $X^{}_{11}$, $X^{}_{12}$ and $X^{}_{13}$ are all real. It is obvious that
the texture of $M^{}_{\rm D}$ is exactly analogous to the pattern of $U$ in
such a specific $\mu$-$\tau$ reflection symmetry scenario.

{\bf Scenario B:} ${\cal T} = {\cal P}$. This interesting possibility has been
discussed for a concrete seesaw model based on the $\mu$-$\tau$ reflection
symmetry (see, e.g., Refs.~\cite{Mohapatra:2015gwa,Xing:2019edp}).
Needless to say, the flavor textures of $M^{}_{\rm D}$ and $M^{}_{\rm R}$ can be
well constrained in this case:
\begin{eqnarray}
M^{}_{\rm D} \hspace{-0.2cm} & = & \hspace{-0.2cm}
\left(\begin{matrix} X^{}_{11} & X^{}_{12} & X^*_{12} \cr
X^{}_{21} & X^{}_{22} & X^{}_{23} \cr
X^*_{21} & X^*_{23} & X^*_{22} \end{matrix}\right) \; ,
\nonumber \\
M^{}_{\rm R} \hspace{-0.2cm} & = & \hspace{-0.2cm}
\left(\begin{matrix} Z^{}_{11} & \hspace{0.08cm} Z^{}_{12} \hspace{0.08cm}
& Z^*_{12} \cr Z^{}_{12} & Z^{}_{22} & Z^{}_{23} \cr
Z^*_{12} & Z^{}_{23} & Z^*_{22} \end{matrix}\right) \; ,
\label{2.16}
%     (2.16)
\end{eqnarray}
where $X^{}_{11}$, $Z^{}_{11}$ and $Z^{}_{23}$ are real. We find that
the number of real free parameters of $M^{}_{\rm D}$ is reduced by half, from
eighteen to nine; and that of $M^{}_{\rm R}$ is also reduced by half, from twelve
to six.

It has been shown in Ref.~~\cite{Mohapatra:2015gwa} that the observed baryon
number asymmetry of the Universe can be interpreted by combining Scenario B
with the leptogenesis mechanism~\cite{Fukugita:1986hr}. Some other forms of
$\cal T$, which respect the $\rm S^{}_3$ symmetry, have also been discussed
in Ref.~\cite{Xing:2019edp} to constrain the seesaw flavor textures and in turn
the validity of leptogenesis. Instead of assuming a special form of $\cal T$
as done in the literature, here we have derived the constraint conditions for
$M^{}_{\rm D}$ and $M^{}_{\rm R}$ in Eq.~(\ref{2.14}) with the help of only the
data-inspired conjecture $U = {\cal P} U^*$ in the canonical seesaw framework.
So our result is certainly more generic and thus could open up some more
possibilities of model building along this line of thought.

After Eq.~(\ref{2.14}) is substituted into Eq.~(\ref{2.2}), the overall neutrino 
mass term ${\cal L}^\prime_\nu$ becomes
\begin{eqnarray}
-{\cal L}^{\prime}_\nu \hspace{-0.2cm} & = & \hspace{-0.2cm}
\frac{1}{2} \hspace{0.05cm} \overline{\left[
{\cal P} \nu^{}_{\rm L} \hspace{0.2cm} {\cal T}^* (N^{}_{\rm R})^c\right]}
\left(\begin{matrix} {\bf 0} & \hspace{0.04cm} M^{*}_{\rm D} \cr
M^\dagger_{\rm D} & \hspace{0.04cm} M^{*}_{\rm R} \end{matrix}\right)
\left[\begin{matrix} \hspace{0.17cm} {\cal P} (\nu^{}_{\rm L})^c
\hspace{0.17cm} \cr {\cal T} N^{}_{\rm R} \end{matrix}
\right] + {\rm h.c.}
\nonumber \\
\hspace{-0.2cm} & = & \hspace{-0.2cm}
\frac{1}{2} \hspace{0.05cm} \overline{\left[
{\cal P} (\nu^{}_{\rm L})^c \hspace{0.37cm} {\cal T} N^{}_{\rm R}\right]}
\left(\begin{matrix} {\bf 0} & M^{}_{\rm D} \cr
M^T_{\rm D} & M^{}_{\rm R} \end{matrix}\right)
\left[\begin{matrix} {\cal P} \nu^{}_{\rm L} \cr
{\cal T}^* (N^{}_{\rm R})^c \end{matrix}
\right] + {\rm h.c.} \; ,
\label{2.17}
%     (2.17)
\end{eqnarray}
in which the apparent mass term of the second equation comes from the Hermitian
conjugation term of the first equation. Comparing between Eq.~(\ref{2.2}) and 
Eq.~(\ref{2.17}), we easily find that ${\cal L}^\prime_\nu$ keeps unchanged 
under the transformations
\begin{eqnarray}
\nu^{}_{\rm L} \to {\cal P} (\nu^{}_{\rm L})^c \; , \quad
N^{}_{\rm R} \to {\cal T}^* (N^{}_{\rm R})^c \; .
\label{2.18}
%     (2.18)
\end{eqnarray}
That is, the neutrino mass term ${\cal L}^\prime_\nu$ is invariant when the
left-handed neutrino fields transform as
\begin{eqnarray}
\left(\begin{matrix} \nu^{}_{e \rm L} \cr \hspace{0.08cm} \nu^{}_{\mu \rm L}
\hspace{0.08cm} \cr \nu^{}_{\tau \rm L} \cr\end{matrix}\right)
\longrightarrow
{\cal P} \left[\begin{matrix} (\nu^{}_{e \rm L})^c \cr
\hspace{0.08cm} (\nu^{}_{\mu \rm L})^c \hspace{0.08cm} \cr
(\nu^{}_{\tau \rm L})^c \cr\end{matrix}\right] =
\left[\begin{matrix} (\nu^{}_{e \rm L})^c \cr (\nu^{}_{\tau \rm L})^c \cr
\hspace{0.08cm} (\nu^{}_{\mu \rm L})^c \hspace{0.08cm} \cr\end{matrix}\right] \; ,
\label{2.19}
%     (2.19)
\end{eqnarray}
while the right-handed neutrino fields transform as in Eq.~(\ref{2.18}). It is well
known that CP would be a good symmetry in the canonical seesaw mechanism
if ${\cal L}^\prime_\nu$  were invariant under the {\it charge conjugation}
and {\it parity} transformations $\nu^{}_{\alpha \rm L} (t, {\bf x}) \to
\big[\nu^{}_{\alpha \rm L} (t, -{\bf x}) \big]^c$ and $N^{}_{\alpha \rm R}
(t, {\bf x}) \to \big[N^{}_{\alpha \rm R} (t, -{\bf x}) \big]^c$
(for $\alpha = e, \mu, \tau$)~\cite{Xing:2011zza}. But the charge-conjugated
interchange between the $(\mu, \tau)$-associated left-handed neutrino fields
in Eq.~(\ref{2.19}), together with an arbitrary unitary CP transformation of the
right-handed neutrino fields, makes CP violation possible even though
${\cal L}^\prime_\nu$ itself keeps invariant in this case. The issue will be
more specific and transparent if a specific form of ${\cal T}$ is taken.
Here let us consider two simple examples as above.

{\bf Scenario A:} ${\cal T} = I$ means that the right-handed neutrinos
have a normal CP transformation,
\begin{eqnarray}
\left(\begin{matrix} N^{}_{e \rm R} \cr N^{}_{\mu \rm R}
\cr N^{}_{\tau \rm R} \cr\end{matrix}\right)
\longrightarrow
I \left[\begin{matrix} (N^{}_{e \rm R})^c \cr
(N^{}_{\mu \rm R})^c \cr
(N^{}_{\tau \rm R})^c \cr\end{matrix}\right] =
\left[\begin{matrix} (N^{}_{e \rm R})^c \cr (N^{}_{\mu \rm R})^c \cr
(N^{}_{\tau \rm R})^c \cr\end{matrix}\right] \; ;
\label{2.20}
%     (2.20)
\end{eqnarray}

{\bf Scenario B:} ${\cal T} = {\cal P}$ yields
\begin{eqnarray}
\left(\begin{matrix} N^{}_{e \rm R} \cr N^{}_{\mu \rm R}
\cr N^{}_{\tau \rm R} \cr\end{matrix}\right)
\longrightarrow
{\cal P} \left[\begin{matrix} (N^{}_{e \rm R})^c \cr
(N^{}_{\mu \rm R})^c \cr
(N^{}_{\tau \rm R})^c \cr\end{matrix}\right] =
\left[\begin{matrix} (N^{}_{e \rm R})^c \cr (N^{}_{\tau \rm R})^c \cr
(N^{}_{\mu \rm R})^c \cr\end{matrix}\right] \; ,
\label{2.21}
%     (2.21)
\end{eqnarray}
where the right-handed neutrino fields transform in the same way as the
left-handed ones in Eq.~(\ref{2.19}). This parallelism between active
and sterile sectors might be helpful for model building. 

Integrating out those heavy degrees of freedom in the canonical seesaw mechanism,
one may arrive at the following effective mass term for three active
neutrinos:
\begin{eqnarray}
-{\cal L}^{}_{\rm mass} = \frac{1}{2} \hspace{0.05cm} \overline{\nu^{}_{\rm L}}
\hspace{0.05cm} M^{}_\nu (\nu^{}_{\rm L})^c + {\rm h.c.} \; ,
\label{2.22}
%     (2.22)
\end{eqnarray}
where
\begin{eqnarray}
M^{}_\nu \simeq - M^{}_{\rm D} M^{-1}_{\rm R} M^T_{\rm D}
\label{2.23}
%     (2.23)
\end{eqnarray}
is known as an approximate seesaw formula for the {\it effective} Majorana neutrino
mass matrix $M^{}_\nu$. In this regard a unitary matrix $U^{}_0$ used to diagonalize
$M^{}_\nu$ just serves as the PMNS matrix (i.e., $U^\dagger_0 M^{}_\nu U^*_0 = D^{}_\nu$).
Substituting Eq.~(\ref{2.14}) into Eq.~(\ref{2.22}) and taking the complex conjugate, 
we obtain $M^{}_\nu = {\cal P} M^*_\nu {\cal P}$ and hence $U^{}_0 = {\cal P} U^*_0$ as a
self-consistent result. It is then easy to check that ${\cal L}^{}_{\rm mass}$
will be unchanged under the transformations
$\nu^{}_{e \rm L} \to (\nu^{}_{e \rm L})^c$,
$\nu^{}_{\mu \rm L} \to (\nu^{}_{\tau \rm L})^c$ and
$\nu^{}_{\tau \rm L} \to (\nu^{}_{\mu \rm L})^c$. Namely, the flavor texture of
$M^{}_\nu$ can be constrained as~\cite{Babu:2002dz,Ma:2002ge,Grimus:2003yn,
Xing:2015fdg}:
\begin{eqnarray}
M^{}_\nu = \left(\begin{matrix} \langle m\rangle^{}_{ee} &
\langle m\rangle^{}_{e \mu} & \langle m\rangle^*_{e \mu} \cr
\langle m\rangle^{}_{e \mu} & \langle m\rangle^{}_{\mu \mu} &
\langle m\rangle^{}_{\mu \tau} \cr
\langle m\rangle^{*}_{e \mu} & \langle m\rangle^{}_{\mu \tau} &
\langle m\rangle^*_{\mu \mu} \end{matrix}\right) \;
\label{2.24}
%     (2.24)
\end{eqnarray}
with $\langle m\rangle^{}_{ee} = \langle m\rangle^{*}_{ee}$
and $\langle m\rangle^{}_{\mu \tau} = \langle m\rangle^{*}_{\mu \tau}$ being real.
Such a minimal flavor symmetry has commonly been referred to as the $\mu$-$\tau$
reflection symmetry~\cite{Harrison:2002et}. Here we have proved possible
existence and correctness of its {\it generalized} version by taking account of
the {\it non-unitarity} of the PMNS matrix $U$ and combining the equalities
$\big|U^{}_{\mu i}\big| = \big|U^{}_{\tau i}\big|$ with the canonical seesaw
mechanism.

Of course, the equalities $\big|U^{}_{\mu i}\big| = \big|U^{}_{\tau i}\big|$
are most likely to hold as a good approximation at low energies. This issue
will become clear after the parameters of flavor mixing and CP violation
are measured to a much better degree of precision in the upcoming experiments
of neutrino oscillations. So it is more reasonable to conjecture
that the equalities $\big|U^{}_{\mu i}\big| = \big|U^{}_{\tau i}\big|$ are
exact and originate from the $\mu$-$\tau$ reflection symmetry at the seesaw
scale, and they become approximate at the electroweak scale as a natural
consequence of the renormalization-group-equation (RGE) running effects (see, e.g.,
Refs.~\cite{Xing:2015fdg,Luo:2014upa,Zhou:2014sya,Huang:2018wqp,Nath:2018zoi,
Huang:2020kgt,Zhao:2020bzx} for some detailed analyses and discussions).

\subsection{An Euler-like parametrization}

In the canonical seesaw framework a full Euler-like parametrization of $U$, $R$,
$S$ and $Q$ in terms of fifteen rotation angles and fifteen phase angles has been 
proposed in Refs.~\cite{Xing:2007zj,Xing:2011ur} with a kind of decomposition like
\begin{eqnarray}
\left( \begin{matrix} U & R \cr S & Q \cr \end{matrix} \right) =
\left( \begin{matrix} P^{}_l & {\bf 0} \cr {\bf 0} & U^\prime_0 \cr \end{matrix}
\right) \left ( \begin{matrix} A & \tilde{R} \cr R^\prime & B \cr \end{matrix} \right )
\left( \begin{matrix} U^{}_0 & {\bf 0} \cr {\bf 0} & I \cr \end{matrix} \right)
= \left( \begin{matrix} P^{}_l A U^{}_0 & P^{}_l \tilde{R} \cr
U^\prime_0 R^\prime U^{}_0 & U^\prime_0 B \cr \end{matrix} \right) \; ,
\label{2.25}
%     (2.25)
\end{eqnarray}
where $P^{}_l = {\rm Diag}\{e^{{\rm i} \varphi^{}_e} , e^{{\rm i} \varphi^{}_\mu} ,
e^{{\rm i} \varphi^{}_\tau}\}$ is an arbitrary phase matrix associated with three
charged-lepton fields, $U^{}_0$ and $U^\prime_0$ are the unitary matrices which
describe the respective primary flavor mixing effects of active and sterile neutrinos,
and $A$ (or $B$) characterizes a slight deviation of $U = P^{}_l A U^{}_0$
(or $Q = U^\prime_0 B$) from $U^{}_0$ (or $U^\prime_0$). But here we are only interested
in $U$ and $R$ that appear in the weak charged-current interactions of massive neutrinos
as shown in Eq.~(\ref{2.6}). To be explicit~\cite{Xing:2011ur},
\begin{eqnarray}
U^{}_0 = \left( \begin{matrix} c^{}_{12} c^{}_{13} & \hat{s}^*_{12}
c^{}_{13} & \hat{s}^*_{13} \cr
-\hat{s}^{}_{12} c^{}_{23} -
c^{}_{12} \hat{s}^{}_{13} \hat{s}^*_{23} & c^{}_{12} c^{}_{23} -
\hat{s}^*_{12} \hat{s}^{}_{13} \hat{s}^*_{23} & c^{}_{13}
\hat{s}^*_{23} \cr
\hat{s}^{}_{12} \hat{s}^{}_{23} - c^{}_{12}
\hat{s}^{}_{13} c^{}_{23} & -c^{}_{12} \hat{s}^{}_{23} -
\hat{s}^*_{12} \hat{s}^{}_{13} c^{}_{23} & c^{}_{13} c^{}_{23}
\cr \end{matrix} \right) \; ,
\label{2.26}
%     (2.26)
\end{eqnarray}
and
\begin{eqnarray}
A \hspace{-0.2cm} & = & \hspace{-0.2cm}
\left( \begin{matrix} c^{}_{14} c^{}_{15} c^{}_{16} & 0 & 0 \cr \vspace{-0.35cm} \cr
\begin{array}{l} -c^{}_{14} c^{}_{15} \hat{s}^{}_{16} \hat{s}^*_{26} -
c^{}_{14} \hat{s}^{}_{15} \hat{s}^*_{25} c^{}_{26} \\
-\hat{s}^{}_{14} \hat{s}^*_{24} c^{}_{25} c^{}_{26} \end{array} &
c^{}_{24} c^{}_{25} c^{}_{26} & 0 \cr \vspace{-0.35cm} \cr
\begin{array}{l} -c^{}_{14} c^{}_{15} \hat{s}^{}_{16} c^{}_{26} \hat{s}^*_{36}
+ c^{}_{14} \hat{s}^{}_{15} \hat{s}^*_{25} \hat{s}^{}_{26} \hat{s}^*_{36} \\
- c^{}_{14} \hat{s}^{}_{15} c^{}_{25} \hat{s}^*_{35} c^{}_{36} +
\hat{s}^{}_{14} \hat{s}^*_{24} c^{}_{25} \hat{s}^{}_{26}
\hat{s}^*_{36} \\
+ \hat{s}^{}_{14} \hat{s}^*_{24} \hat{s}^{}_{25} \hat{s}^*_{35}
c^{}_{36} - \hat{s}^{}_{14} c^{}_{24} \hat{s}^*_{34} c^{}_{35}
c^{}_{36} \end{array} &
\begin{array}{l} -c^{}_{24} c^{}_{25} \hat{s}^{}_{26} \hat{s}^*_{36} -
c^{}_{24} \hat{s}^{}_{25} \hat{s}^*_{35} c^{}_{36} \\
-\hat{s}^{}_{24} \hat{s}^*_{34} c^{}_{35} c^{}_{36} \end{array} &
c^{}_{34} c^{}_{35} c^{}_{36} \cr \end{matrix} \right) \; , \hspace{0.6cm}
\nonumber \\
\tilde{R} \hspace{-0.2cm} & = & \hspace{-0.2cm}
\left( \begin{matrix} \hat{s}^*_{14} c^{}_{15} c^{}_{16} &
\hat{s}^*_{15} c^{}_{16} & \hat{s}^*_{16} \cr \vspace{-0.35cm} \cr
\begin{array}{l} -\hat{s}^*_{14} c^{}_{15} \hat{s}^{}_{16} \hat{s}^*_{26} -
\hat{s}^*_{14} \hat{s}^{}_{15} \hat{s}^*_{25} c^{}_{26} \\
+ c^{}_{14} \hat{s}^*_{24} c^{}_{25} c^{}_{26} \end{array} & -
\hat{s}^*_{15} \hat{s}^{}_{16} \hat{s}^*_{26} + c^{}_{15}
\hat{s}^*_{25} c^{}_{26} & c^{}_{16} \hat{s}^*_{26} \cr \vspace{-0.35cm} \cr
\begin{array}{l} -\hat{s}^*_{14} c^{}_{15} \hat{s}^{}_{16} c^{}_{26}
\hat{s}^*_{36} + \hat{s}^*_{14} \hat{s}^{}_{15} \hat{s}^*_{25}
\hat{s}^{}_{26} \hat{s}^*_{36} \\ - \hat{s}^*_{14} \hat{s}^{}_{15}
c^{}_{25} \hat{s}^*_{35} c^{}_{36} - c^{}_{14} \hat{s}^*_{24}
c^{}_{25} \hat{s}^{}_{26}
\hat{s}^*_{36} \\
- c^{}_{14} \hat{s}^*_{24} \hat{s}^{}_{25} \hat{s}^*_{35}
c^{}_{36} + c^{}_{14} c^{}_{24} \hat{s}^*_{34} c^{}_{35} c^{}_{36}
\end{array} &
\begin{array}{l} -\hat{s}^*_{15} \hat{s}^{}_{16} c^{}_{26} \hat{s}^*_{36}
- c^{}_{15} \hat{s}^*_{25} \hat{s}^{}_{26} \hat{s}^*_{36} \\
+c^{}_{15} c^{}_{25} \hat{s}^*_{35} c^{}_{36} \end{array} &
c^{}_{16} c^{}_{26} \hat{s}^*_{36} \cr \end{matrix} \right) \; ,
\label{2.27}
%     (2.27)
\end{eqnarray}
where $c^{}_{ij} \equiv \cos\theta^{}_{ij}$, $s^{}_{ij} \equiv
\sin\theta^{}_{ij}$ and $\hat{s}^{}_{ij} \equiv s^{}_{ij} \hspace{0.05cm}
e^{{\rm i}\delta^{}_{ij}}$ with $\theta^{}_{ij}$ lying in the first quadrant
(for $ij = 12$, $13$, $\cdots$). It is clear that
the triangular matrix $A$ characterizes to what extent the PMNS matrix $U$
may deviate from $U^{}_0$. Since the interplay between active and sterile
neutrino sectors is expected to be weak enough, it is rather
reasonable to make the following approximations to $A$ and
$\tilde{R}$~\cite{Xing:2011ur}:
\begin{eqnarray}
A \hspace{-0.2cm} & \simeq & \hspace{-0.2cm}
I - \frac{1}{2} \left( \begin{matrix} s^2_{14} +
s^2_{15} + s^2_{16} & 0 & 0 \cr
2\hat{s}^{}_{14}
\hat{s}^*_{24} + 2\hat{s}^{}_{15} \hat{s}^*_{25} + 2\hat{s}^{}_{16}
\hat{s}^*_{26} & s^2_{24} + s^2_{25} + s^2_{26}
& 0 \cr
2\hat{s}^{}_{14} \hat{s}^*_{34} + 2\hat{s}^{}_{15}
\hat{s}^*_{35} + 2\hat{s}^{}_{16} \hat{s}^*_{36} & 2\hat{s}^{}_{24}
\hat{s}^*_{34} + 2\hat{s}^{}_{25} \hat{s}^*_{35} + 2\hat{s}^{}_{26}
\hat{s}^*_{36} & s^2_{34} + s^2_{35} + s^2_{36} \cr \end{matrix}
\right) + {\cal O}(s^4_{ij}) \; ,
\nonumber \\
\tilde{R} \hspace{-0.2cm} & \simeq & \hspace{-0.2cm}
\left( \begin{matrix} \hat{s}^*_{14} &
\hat{s}^*_{15} & \hat{s}^*_{16} \cr
\hat{s}^*_{24} & \hat{s}^*_{25} &
\hat{s}^*_{26} \cr
\hat{s}^*_{34} & \hat{s}^*_{35} &
\hat{s}^*_{36} \cr \end{matrix} \right) + {\cal O}(s^3_{ij}) \; ,
\label{2.28}
%     (2.28)
\end{eqnarray}
where all the $s^{}_{ij}$ terms (for $i = 1, 2, 3$ and $j = 4, 5, 6$) are
expected to be below or even far below ${\cal O}(10^{-1})$. So $U$ can be
treated as a unitary matrix to a good degree of accuracy (i.e., $A \simeq I$
and $U \simeq P^{}_l U^{}_0$ up to the ${\cal O}(s^2_{ij})$ level). In this case
combining Eqs.~(\ref{2.26}) and (\ref{2.28}) with $U = {\cal P} U^*$ and
$R = {\cal P} R^*$ allows us to obtain some approximate constraint conditions:
\begin{eqnarray}
e^{{\rm i} 2\varphi^{}_e} \simeq e^{{\rm i} 2\left(\varphi^{}_e - \delta^{}_{12}\right)}
\simeq e^{{\rm i} 2\left(\varphi^{}_e - \delta^{}_{13}\right)} \simeq 1 \;
\label{2.29}
%     (2.29)
\end{eqnarray}
from $U^{}_{e i} \simeq U^*_{e i}$ (for $i = 1, 2, 3$), and
\begin{eqnarray}
\left(s^{}_{12} c^{}_{23} + c^{}_{12} s^{}_{13} s^{}_{23} e^{{\rm i} \delta}\right)
e^{{\rm i}\left(\varphi^{}_\mu + \varphi^{}_\tau + 2\delta^{}_{12} + \delta^{}_{23}\right)}
\hspace{-0.2cm} & \simeq & \hspace{-0.2cm}
- \left(s^{}_{12} s^{}_{23} - c^{}_{12} s^{}_{13} c^{}_{23} e^{-{\rm i}\delta}\right) \; ,
\nonumber \\
\left(c^{}_{12} c^{}_{23} - s^{}_{12} s^{}_{13} s^{}_{23} e^{{\rm i} \delta}\right)
e^{{\rm i}\left(\varphi^{}_\mu + \varphi^{}_\tau + \delta^{}_{23}\right)}
\hspace{-0.2cm} & \simeq & \hspace{-0.2cm}
- \left(c^{}_{12} s^{}_{23} + s^{}_{12} s^{}_{13} c^{}_{23} e^{-{\rm i}\delta}\right) \; ,
\hspace{1cm}
\nonumber \\
s^{}_{23} e^{{\rm i}\left(\varphi^{}_\mu + \varphi^{}_\tau - \delta^{}_{23}\right)}
\hspace{-0.2cm} & \simeq & \hspace{-0.2cm} c^{}_{23} \;
\label{2.30}
%     (2.30)
\end{eqnarray}
from $U^{}_{\mu i } \simeq U^*_{\tau i}$ (for $i = 1, 2, 3$), where
$\delta \equiv \delta^{}_{13} - \delta^{}_{12} - \delta^{}_{23}$ has been defined;
as well as
\begin{eqnarray}
e^{{\rm i} 2\left(\varphi^{}_e - \delta^{}_{14}\right)} \simeq
e^{{\rm i} 2\left(\varphi^{}_e - \delta^{}_{15}\right)}
\simeq e^{{\rm i} 2\left(\varphi^{}_e - \delta^{}_{16}\right)} \simeq 1 \;
\label{2.31}
%     (2.31)
\end{eqnarray}
from $R^{}_{e i} \simeq R^*_{e i}$ (for $i = 1, 2, 3$), and
\begin{eqnarray}
s^{}_{24} e^{{\rm i}\left(\varphi^{}_\mu + \varphi^{}_\tau - \delta^{}_{24}
- \delta^{}_{34}\right)} \hspace{-0.2cm} & \simeq & \hspace{-0.2cm}
s^{}_{34} \; ,
\nonumber \\
s^{}_{25} e^{{\rm i}\left(\varphi^{}_\mu + \varphi^{}_\tau - \delta^{}_{25}
- \delta^{}_{35}\right)} \hspace{-0.2cm} & \simeq & \hspace{-0.2cm}
s^{}_{35} \; ,
\nonumber \\
s^{}_{26} e^{{\rm i}\left(\varphi^{}_\mu + \varphi^{}_\tau - \delta^{}_{26}
- \delta^{}_{36}\right)} \hspace{-0.2cm} & \simeq & \hspace{-0.2cm}
s^{}_{36} \; 
\label{2.32}
%     (2.32)
\end{eqnarray}
from $R^{}_{\mu i } \simeq R^*_{\tau i}$ (for $i = 1, 2, 3$). As a consequence, we arrive at
%%%%%%%%%%%%%%%%%%%%%%%%%%%%%%%%%%%%%%%%%%%%%%%%%%%%%%%%%%%%%%%%%%%%%%%%%%%%%%%%%%%%%%%
\footnote{Note that the phase convention of $U^{}_0$ used in Eq.~(\ref{2.26}) is somewhat
different from those used by some other authors (see, e.g.,
Refs.~\cite{ParticleDataGroup:2020ssz,Xing:2019vks,Zhou:2014sya}). Of course, it is trivial
to establish a direct relationship between any two kinds of phase conventions for the
PMNS matrix when it is assumed to be exactly unitary.}
%%%%%%%%%%%%%%%%%%%%%%%%%%%%%%%%%%%%%%%%%%%%%%%%%%%%%%%%%%%%%%%%%%%%%%%%%%%%%%%%%%%%%%%
\begin{eqnarray}
\theta^{}_{23} \simeq \frac{\pi}{4} \; , \quad \delta \simeq \pm\frac{\pi}{2} \; , \quad
\delta^{}_{12} \simeq 0 ~{\rm or}~ \pi \; , \quad
\delta^{}_{13} \simeq 0 ~{\rm or}~ \pi \; , \quad
\delta^{}_{23} \simeq \pm \frac{\pi}{2} \; ,
\label{2.33}
%     (2.33)
\end{eqnarray}
together with $\varphi^{}_e \simeq 0$ or $\pi$ and
$\varphi^{}_\mu + \varphi^{}_\tau \simeq \delta^{}_{23} \simeq \pm \pi/2$;
and in the same time,
\begin{eqnarray}
\theta^{}_{2j} \simeq \theta^{}_{3j} \; , \quad
\delta^{}_{1j} \simeq 0 ~{\rm or}~ \pi \; , \quad
\delta^{}_{2j} + \delta^{}_{3j} \simeq
\varphi^{}_\mu + \varphi^{}_\tau \simeq \pm\frac{\pi}{2} \; ,
\label{2.34}
%     (2.34)
\end{eqnarray}
where $j = 4, 5, 6$. Note that $\varphi^{}_e$, $\varphi^{}_\mu$ and
$\varphi^{}_\tau$ are not observable in any realistic experiments,
but they should not be ignored when using the $\mu$-$\tau$ reflection
symmetry to constrain those observable flavor mixing angles and
CP-violating phases as discussed above.

So far a lot of attention has been paid to $\theta^{}_{23} \simeq \pi/4$
and $\delta \simeq \pm\pi/2$ obtained in Eq.~(\ref{2.33}) with the help of the
$\mu$-$\tau$ reflection symmetry of
${\cal L}^{}_{\rm mass}$, either from the viewpoint of model building or
in the in-depth studies of neutrino phenomenology~\cite{Xing:2015fdg}.
When applying such a minimal flavor symmetry to the seesaw mechanism, one
should carefully examine whether its generalized version advocated in the
present work should be taken into account or not. In particular, the correlation
between $U$ and $R$ via the exact seesaw formula and unitarity conditions
are sensitive to some lepton-number-violating processes (e.g., the neutrinoless
double-beta decays~\cite{Xing:2009ce} and leptogenesis~\cite{Fukugita:1986hr})
and lepton-flavor-violating processes (e.g., $\mu^- \to e^- + \gamma$)
in which the interplay between active and sterile Majorana neutrinos
cannot be neglected~\cite{Xing:2019edp,Xing:2020ivm,Zhang:2021tsq,Zhang:2021jdf,
Zhao:2021dwc}.

At this point it is worth mentioning that a generalized $\mu$-$\tau$ reflection
symmetry is also applicable to the {\it minimal} seesaw model which contains
only two species of right-handed neutrinos~\cite{Frampton:2002qc}. In this
simpler case the overall neutrino mass term keeps invariant when the
three left-handed neutrino fields transform as
$\nu^{}_{e \rm L} \to (\nu^{}_{e \rm L})^c$,
$\nu^{}_{\mu \rm L} \to (\nu^{}_{\tau \rm L})^c$,
$\nu^{}_{\tau \rm L} \to (\nu^{}_{\mu \rm L})^c$ and the
two right-handed neutrino fields undergo an arbitrary unitary
CP transformation (see, e.g., Refs.~\cite{Nath:2018hjx,Zhao:2021tgi,Xing:2020ald}
for a few concrete models of this kind), and the 
approximate relations obtained in Eqs.~(\ref{2.31}), (\ref{2.32}) and (\ref{2.34})
keep valid only if the active-sterile flavor mixing parameters
$\theta^{}_{i 6}$ and $\delta^{}_{i 6}$ (for $i = 1, 2, 3$) are switched off.
Needless to say, such a simplified version of the canonical seesaw
mechanism is expected to be much more predictive and thus can be more
easily tested after it is combined with the $\mu$-$\tau$ reflection 
symmetry in a consistent manner.

\section{Summary}

In the canonical seesaw framework we have identified a minimal flavor symmetry
lying behind the equalities
$\big|U^{}_{\mu i}\big| = \big|U^{}_{\tau i}\big|$ (for $i = 1, 2, 3$)
that are strongly favored by current experimental data on neutrino oscillations.
In view of the fact that the PMNS matrix $U$ is correlated with the $3\times 3$
flavor mixing matrix $R$ characterizing the strength of weak charged-current
interactions of heavy Majorana neutrinos via the exact seesaw formula and the
unitarity condition $UU^\dagger + RR^\dagger = I$, we have shown that
$\big|U^{}_{\mu i}\big| = \big|U^{}_{\tau i}\big|$ can automatically lead to
$\big|R^{}_{\mu i}| = |R^{}_{\tau i}\big|$ (for $i = 1, 2, 3$).
We have further proved that behind these two sets of equalities and the T2K
evidence for leptonic CP violation is a very simple flavor symmetry --- the
overall neutrino mass term keeps invariant when the left-handed neutrino fields
transform as $\nu^{}_{e \rm L} \to (\nu^{}_{e \rm L})^c$,
$\nu^{}_{\mu \rm L} \to (\nu^{}_{\tau \rm L})^c$,
$\nu^{}_{\tau \rm L} \to (\nu^{}_{\mu \rm L})^c$ and the right-handed neutrino
fields undergo an arbitrary unitary CP transformation.
Such a generalized $\mu$-$\tau$ reflection symmetry is expected to help a lot to
constrain the flavor textures of active and sterile neutrinos and hence
enhance predictability and testability of the seesaw mechanism.
We have illustrated how the flavor mixing angles and CP-violating phases are
explicitly constrained by making use of a full Euler-like parametrization of
the PMNS matrix $U$ and its counterpart $R$.

We must stress that the canonical seesaw mechanism may serve as the most natural
and popular theoretical framework for the origin of tiny neutrino
masses at a very low cost for going beyond the SM. So a possible minimal flavor
symmetry underlying this mechanism, like the generalized $\mu$-$\tau$
reflection symmetry that we have explored in this work, certainly
deserves our serious attention. A further and comprehensive study of such a
flavor symmetry, including how to naturally break it, how to naturally embed 
it into a viable gauge model and how to link it with various measurements of
lepton-number-violating and lepton-flavor-violating processes, is well worth 
the wait.

\section*{Acknowledgements}

The author is greatly indebted to Di Zhang for many useful discussions
and clarifying several important issues of this work. This research has been
supported in part by the National Natural Science Foundation of
China under grant No. 12075254 and grant No. 11835013.

\end{document}